\documentclass[12pt]{iopart}
  \expandafter\let\csname equation*\endcsname\relax
  \expandafter\let\csname endequation*\endcsname\relax
\usepackage{amsmath, amssymb}
\usepackage{graphicx}
\usepackage{xcolor}
\usepackage{url}
\usepackage{amsfonts}
\newsavebox{\myhbar}
\savebox{\myhbar}{$\hbar$}
\usepackage{xcolor}
\usepackage{cite}
\usepackage{enumerate}
\graphicspath{ {./images/} }
\numberwithin{equation}{section}
\renewcommand*{\hbar}{\mathalpha{\usebox{\myhbar}}}
\newcommand{\bey}{\begin{eqnarray}}
\newcommand{\eey}{\end{eqnarray}}

\begin{document}

\title{Quantum Superposition of Two Gravitational Cat States}
\author{C. Anastopoulos$^1$ and B. L. Hu$^2$}

\address{$^1$Department of Physics, University of Patras, 26500 Patras, Greece.}

\address{$^2$Maryland Center for Fundamental Physics and Joint Quantum Institute,\\ University of
Maryland, College Park, Maryland 20742-4111 U.S.A.}

\ead{anastop@upatras.gr,blhu@umd.edu}

\date{\today}

\begin{abstract}
We extend our earlier work \cite{AnHu15} on probing a gravitational cat state (gravcat) to the quantum superposition of two gravcats in an exemplary model and in Bose-Einstein condensates (BEC). In addition to its basic theoretical values in gravitational quantum physics and macroscopic quantum phenomena, this investigation can provide some theoretical support to  experimental proposals for measuring gravity-induced entanglement and the quantum nature of perturbative gravity. In the first part we consider cat  states generated by double-well potentials. A pair of gravcats, each approximated as a two-level system,  is characterized by \textit{gravity-induced Rabi oscillations}, and by \textit{gravity-induced entanglement} of its energy eigenstates. In the second part  we turn to a (non-relativistic) quantum field theory description and derive a \textit{gravitational Gross-Pitaevsky equation} for gravcats formed in Bose-Einstein condensates. Using a mathematical analogy to quantum rotors, we explore the properties of the two-gravcat system for BECs, its physical consequences and observational possibilities. Finally we discuss our results in comparison to predictions of alternative quantum theories, and we explain their implications.
\end{abstract}

\section{Introduction}

In a quantum description of  matter  a  single  motionless massive particle can in principle be in a superposition state of  two spatially-separated locations, i.e., a Schr\"odinger cat state. We use the term gravitational cat (gravcat) to refer  to such states for objects that gravitate \cite{AnHu15}. Understanding the behavior of such states is of foundational interest, especially in the fields of  gravitational quantum physics \cite{GQP} and macroscopic quantum phenomena \cite{MQP}; for samples of mechanical, non-BEC types, see, e.g.,\cite{MechMQP}).

%It is also relevant  to proposals of tabletop experiments which purport to showing quantum gravity \cite{Bose17} or the quantum nature of gravity \cite{Vedral17}.

\subsection{Motif: Gravitation of quantum matter}

The main theoretical motivation behind this work is the need to understand how quantum systems interact gravitationally. Contrasting the abundance of theories for quantum gravity purporting to describe physics at the Planck scale, at today's low energy scale even though we  have well-tested theories for gravitation and for quantum matter, experimental tests or proposals about their interaction are miserably scarce, even in the accessible non-relativistic (NR), weak-gravitational (WG) field regimes. The famous Colella-Overhauser-Werner (COW) experiment \cite{COW} established that, for non-relativistic particles, the effects of a background gravitational field are accounted by the addition of a potential term in the Hamiltonian operator. This was confirmed by experiments on neutrons bouncing off a horizontal mirror \cite{nbounce}, which demonstrated the existence of bound states due to gravitational field.

Experiments testing how quantum systems gravitate include determining (i) the gravitational force generated by a quantum distribution of matter, and (ii) the gravitational interaction between two different quantum matter distributions. In classical physics, gravity at the non-relativistic, weak-field limit is non-dynamical. It is described solely by the gravitational potential, which is completely `slaved' to the mass density through Poisson's equation. By `slaved' we mean the potential is fully determined by the theory's first-class constraints \cite{constraints}, and it is not an independent physical observable.

The most natural hypothesis is that this property of classical gravity holds when quantum matters interact in the presence of gravity\footnote{This is the working principle at the semiclassical gravity level of description, but note the fundamental differences in the  so-called M\/oller-Rosenfeld equation \cite{MolRos} which many alternative quantum theories \cite{BassiRMP} refer to as semiclassical gravity \cite{HuAQT} and use in  the derivation of the Newton-Schr\"odinger (NS) equation \cite{NS}, versus semiclassical gravity based on the semiclassical Einstein equation derived from the Einstein equation of classical general relativity (GR) with self-consistent backreaction of a regularized energy-momentum tensor of the quantum matter field (QFT), or quantum field theory in the large N limit  \cite{HuVer20}. NS equations derived without assuming a large N limit do not originate from GR+QFT \cite{AHNS}.    Further discussions of the implications of the pure gauge property of Newtonian potential on the quantum nature of gravity can be found in the Sec. 5.1.1.}, i.e., the gravitational potential is an auxiliary operator defined as a function of an appropriately regularized mass density operator $\hat{\mu}({\pmb x})$. The potential induces a non-local term in the Hamiltonian of the form
 \bey
 \hat{V} = - G \int dx dx' \frac{\hat{\mu}({\pmb x}) \hat{\mu}({\pmb x}')}{|{\pmb x} - {\pmb x}'|}, \label{Coul}
 \eey
 where $G$ is Newton's constant.
This result is derived   by quantising the weak-gravity, non-relativistic limit of General Relativity (GR) interacting with classical matter   \cite{AHNSE}. Alternatively, we can take the non-relativistic limit after quantization of the matter-gravity system, i.e., in an effective quantum field theory (QFT) that describes the interaction of quantum matter with linearized gravity \cite{AnHu13}. The first derivation is more general: the true degrees of freedom of gravity completely separate from the gravitational potential, hence, we need not assume that they are quantized--- for thought experiments suggesting that their quantization is necessary, or otherwise see, e.g., \cite{PagGei,EppHan,Carlip,Baym,BWGCBA}.  The only assumption is that the usual quantization rules apply for matter in the presence of gravity in the weak-field, non-relativistic  (Newtonian) regime.

For this reason, the assumption that the Newtonian potential is slaved to quantum matter is the most conservative. It implies that  quantum superpositions of macroscopically distinct states (cat states) generate quantum superpositions of the gravitational force, which are in principle measurable \cite{DAH}, but says nothing about the quantum nature of gravity.  The present  work adopts this theoretical framework: we analyse the gravitational interaction of a pair of cat states (gravcats), and identify their characteristic behaviors.

The detection of gravity-induced entanglement is discussed in recently proposed tabletop experiments, e.g., \cite{Bose17, Vedral17}. These authors claim that observation of this effect demonstrates quantum gravity \cite{Bose17} or the quantum nature of gravity \cite{Vedral17}. We are pleased with their proposed experiments, but we disagree with their claims \cite{AnHucr}: We maintain that  experiments involving only the Newtonian potential are not sufficient for the demonstration of quantized gravity, certainly not quantum gravity proper at the Planck energy scale, but also not even quantized perturbative gravity at today's low energy scale. The reason is that such experiments do not involve the  dynamical propagating degrees of freedom (the transverse-traceless perturbations) -- only the detection of gravitons \cite{Dyson} can prove the quantum nature of perturbative gravity. We shall return to this important point in Sec. 5.

There are alternative theoretical constructs. Penrose has advanced an influential argument that superpositions of states with macroscopically distinct mass densities cannot be stable and has provided an estimate for their decoherence rate \cite{Penrose, Penrose2}. The same rate is obtained by Diosi's gravitational decoherence model \cite{Diosi}, in which the (Newtonian) gravitational field generates non-unitary dynamics (in addition to any potential terms in the Hamiltonian). Hence, according to the Diosi-Penrose (DP) model, no gravcats are possible. Conversely, any test of gravcats is also a test of gravitational decoherence of the DP type and  observation of gravcats spells the demise of DP models---see, for example, the discussion in Ref. \cite{Pino,CBPU}.

The DP model is the most influential because it involves essentially no free parameters (at least in the domain of experimental interest). However, there are other models of gravitational decoherence \cite{gravdec}, which lead to different predictions. We take up this issue in Sec. 5.

For a more detailed description of the background, our motivation and the set goals of this line of investigation, please refer to the Introduction of our earlier paper \cite{AnHu15} and references therein.

\subsection{This work: Interacting gravcats}

In the first part of this paper, we consider cat states that correspond to energy eigenstates of a particle in a double-well potential. Each gravcat can be well approximated by a qubit. The gravitational interaction between particles induces a coupling terms between two qubits, leading to
  \textit{gravity-induced Rabi oscillations} for some initial states and to \textit{gravity-induced entanglement} of its energy eigenstates.

In the second part of this paper, we turn to a  (non-relativistic) quantum field theory description of gravcats. This makes evident that the gravitational interaction is generated by the term (\ref{Coul}). Furthermore, it allows for the treatment of multi-particle gravcats, in particular ones that correspond to Bose-Einstein condensates (BECs). The latter can be used for a simulation of the gravity-induced oscillations predicted for a single particle, but they are also of interest by their own merit. We derive a \textit{gravitational Gross-Pitaevsky equation} for the gravitational interaction of BECs.

In the two-mode approximation, a BEC in a double-well potential is mathematically equivalent to a single quantum rotor. Hence a pair of BEC gravcats corresponds to a pair of interacting quantum rotors. We analyse this system and discuss its physical consequences and observational possibilities.

\bigskip

The plan of this paper is the following. In Sec. 2, we model the interaction of two gravcats, each defined through a double-well potential, as an interaction of  a pair of qubits. In Sec. 3, we improve on the qubit approximation for this system, using the semi-classical approximation for the eigenstates of the double-well potential. In Sec. 4, we undertake a QFT analysis of the system, and provide a description of interacting gravcats for BECs. In Sec. 5 we conclude with a discussion of the broader context of our results, their implications about the quantum nature of gravity and  provide constraints to alternative quantum theories.

  \section{Gravitational interaction of qubits}

In this section, we present an elementary models for the interaction of two-gravcats, each corresponding to a qubit.

  Consider a particle of mass $m$ in one dimension with Hamiltonian $\hat{H}_0 = \frac{1}{2m} \hat{p}^2 + U(\hat{x})$. The potential $U(x)$ corresponds to a symmetric double well, with local minima at $x = \pm \frac{1}{2}L$. We assume that $U(x)$ is even, so all eigenstates of $\hat{H}$ are parity definite. The lowest energy eigenstate $|g \rangle$ is always parity symmetric and the first excited $| e \rangle$ is parity antisymmetric. We denote the energy difference between  $| e \rangle$ and $|g \rangle$  by $\omega$.

  We define the states
  \bey
  |\pm\rangle = \frac{1}{2} \left(|g\rangle \pm |e\rangle\right).
  \eey
The Landau-Lifschitz approximation consists in identifying   the states $|\pm \rangle$ with the semi-classical wave functions localized around $x = \pm \frac{1}{2}L$. The overlap between the wave functions $\psi_+(x) := \langle x|+\rangle$ and $\psi_-(x) := \langle x|-\rangle$ is negligible. Hence,  we can approximate the action of the position operator $\hat{x}$ on those states as
  \bey
  \hat{x} |\pm\rangle \simeq \pm \frac{L}{2}|\pm\rangle. \label{xapp}
  \eey

  We will restrict to the two-dimensional subspace spanned by $|g\rangle$ and $|e\rangle$. This is a reasonable approximation, if the system is prepared in an environment of temperature $T$ much smaller than the energy $E_2$ of the next excited state. In general, $\omega << E_2$, as the energy split of the two levels is obtained by the symmetry breaking of the classical $Z_2$ invariance of the Hamiltonian through an exponentially suppressed term---see, Eq. (\ref{omexp})---that corresponds to tunneling between the two wells.

   We choose the energy scale so that the energy of $|g\rangle$ is $- \frac{1}{2} \omega$ and the energy of $|e\rangle$ is $ \frac{1}{2} \omega$. The Hamiltonian then reads $\hat{H} = \frac{1}{2} \omega \hat{\sigma}_z$. By Eq. (\ref{xapp}),
  \bey
  \hat{x} =  \frac{L}{2} \hat{\sigma}_x. \label{xsigma}
  \eey

   \begin{figure}
    \centering

 \includegraphics[width=0.65\textwidth]{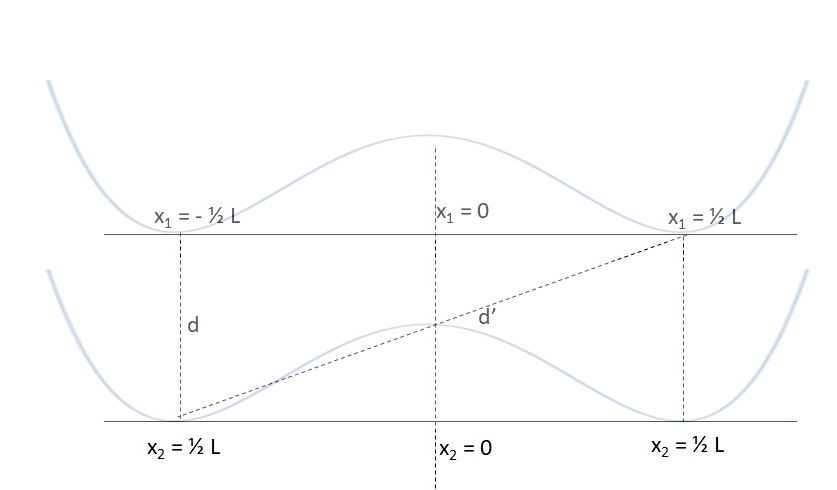}

    \caption{ Geometry of the gravitational  interaction between two Schr\"odinger's cats.}
\end{figure}

Consider now a pair of such systems, in the geometry of Fig. 1. Each double-well lies along a different axis: the two axes are parallel at distance $d$. Taking into
account the gravitational interaction between the two masses,  the total Hamiltonian is
  \bey
  \hat{H} = \hat{H}_0\otimes \hat{I} + \hat{I} \otimes \hat{H}_0 + \hat{V}
  \eey
  where $\hat{V}$ is the Newtonian potential term
  \bey
  \hat{V} = - Gm^2[D(\hat{x}_1, \hat{x}_2)]^{-1} \label{poten}
  \eey
  where $D(x_1, x_2):= \sqrt{d^2+(x_1-x_2)^2}$.

  The operator $\hat{V}$ is diagonal in the position basis. Hence, when the particles are approximated by qubits, $\hat{V}$  is diagonal in the basis that consists of the vectors $|+, +\rangle, |+, - \rangle, |-, +\rangle$ and $|-, - \rangle$. The only non-zero matrix elements of $\hat{V}$ are
  \bey
  \langle +, +|\hat{V} |+,+\rangle =   \langle -, -|\hat{V} |-,- \rangle = - \frac{\alpha}{d} \\
    \langle +, -|\hat{V} |+,-\rangle =   \langle -, +|\hat{V} |-,+ \rangle = - \frac{\alpha}{d'},
  \eey
  where $d' = \sqrt{d^2 +L^2}$, and we wrote $\alpha = Gm^2$.

  It follows that
  \bey
  \hat{V} = - \frac{1}{2} \left( \frac{\alpha}{d} +  \frac{\alpha}{d'}\right) \hat{I} - \frac{1}{2}\left( \frac{\alpha}{d} -  \frac{\alpha}{d'}\right) \hat{\sigma}_x \otimes \hat{\sigma}_x.
  \eey

Ignoring the constant term in $\hat{V}$, the total Hamiltonian of the system is
\bey
\hat{H} = \frac{1}{2}\omega (\hat{I} \otimes \hat{\sigma}_z + \hat{\sigma}_z \otimes \hat{I}) - \mho  \hat{\sigma}_x \otimes \hat{\sigma}_x, \label{qubit}
\eey
  where
  \begin{eqnarray}
  \mho :=\frac{\alpha}{2}(\frac{1}{d} -  \frac{1}{d'}). \label{u}
\end{eqnarray}
  In matrix form,
  \bey
  \hat{H} = \left( \begin{array}{cccc} \omega & -\mho & 0 &0 \\ -\mho &-\omega & 0  & 0 \\ 0 & 0 & 0 & -\mho \\ 0  & 0 & -\mho & 0 \end{array}\right). \label{hh}
  \eey
in the ordered basis $ |e,e\rangle, |g,g\rangle, |e,g \rangle, |g, e\rangle$.

 In increasing order, the eigenvalues of the Hamiltonian are $ - \omega', - \mho, \mho$ and $\omega'$, where  $\omega' = \sqrt{\omega^2+\mho^2}$.
The evolution operator is
\bey
\hspace{-2cm}e^{-i\hat{H}t} = \left( \begin{array}{cccc} \cos (\omega' t)- \frac{i\omega}{\omega'} \sin(\omega't)&  \frac{i\mho }{\omega'} \sin(\omega't)& 0 &0 \\
\frac{i\mho }{\omega'} \sin(\omega't)&\cos (\omega' t)+ \frac{i\omega}{\omega'} \sin(\omega't)&0&0\\
0&0&\cos(\mho t)& i \sin(\mho t) \\ 0&0&i \sin(\mho t) &\cos(\mho t)
\end{array}
\right)
\eey

 The gravitational interaction term causes factorized initial states to evolve towards entangled ones. For example, an initial state $|e, e\rangle$ evolves to
\begin{eqnarray}
|\psi(t)\rangle = [\cos (\omega' t)- \frac{i\omega}{\omega'} \sin(\omega't) ]|e, e\rangle + \frac{i\mho }{\omega'} \sin(\omega't) |g, g\rangle.  \label{eevv}
\end{eqnarray}
 A natural measure of entanglement is the purity $\gamma = 1 - Tr \hat{\rho}_1^2$, of the reduced density matrix $\hat{\rho}_1$ for the first qubit. For Eq. (\ref{eevv}),
 \bey
 \gamma =  \frac{2\mho^2}{\omega^{\prime 2}} \sin^2\omega't (1 - \frac{\mho^2}{\omega^{\prime 2}}\sin^2\omega't).
 \eey
For $\mho << \omega$, $\gamma$ is of order $(\mho/\omega)^2$.

  Furthermore, when the system is in contact with a thermal reservoir, at very low temperatures the asymptotic state is expected to approach the ground state of the system
 \bey
 |0\rangle = \frac{1}{\sqrt{2}}\left( - \sqrt{1 - \frac{\omega}{\omega'}} |e, e \rangle + \sqrt{1 + \frac{\omega}{\omega'}} |g, g \rangle \right), \label{grrr}
 \eey
 which is entangled.
 For $\mho << \omega$, $\gamma = \frac{\mho^2}{2\omega^2}$. The maximum value $\gamma = \frac{1}{2}$ is obtained for   $\omega \rightarrow 0$, and it corresponds to a state of maximal entanglement.

Time evolution is simpler in the subspace spanned by $|e,g \rangle$ and $|g, e \rangle$. For example, if the system is initially prepared in the $|e, g\rangle$ state, the state at time $t$ is
\begin{eqnarray}
|\psi(t)\rangle = \cos (\mho t) |e, g\rangle + i \sin (\mho t) |g,e\rangle.
\end{eqnarray}
Hence, the transfer of energy between the two subsystems  is modulated by  an oscillation of frequency $\mho $ caused by the Newtonian interaction. For $d$ and $L$ of the order of a micron, a period of the order of one minute requires  particles with mass   $m \sim 10^{11} amu$.

 The experimental realization of this system requires that dissipation and noise (decoherence / dephasing) effects can be kept small. Such effects depend crucially on coupling between the two-level system and the environment, the environment's structure and temperature. The full analysis of these phenomena is quite complex \cite{LCDM}, however, an order of magnitude estimate is possible in the weak coupling limit and for an Ohmic environment. Then, the dissipation and relaxation rates are of the order of $\Gamma \sim \alpha \omega \coth\frac{\omega}{2T}$, where $\alpha << 1$ is the dimensionless coupling constant (that depends on the environment properties)and $T$ is the temperature. These rates depend on the particle's mass only indirectly, through $\omega$. At the low temperature limit, $\omega >> T$, $\Gamma \sim  \alpha \omega$, and it is necessary that $\alpha$ be of the order of $\mho/\omega$ or smaller, in order to see Rabi oscillations.

The experiments proposed in Refs. \cite{Bose17, Vedral17, KTPP} emphasize the creation of entanglement as a criterion for identifying gravitational effects in quantum systems.  This may not be the most efficient criterion. Entanglement is typically quadratic in the small parameter $\mho/\omega $. In contrast, the frequency of Rabi oscillations is proportional to $\mho $. It seems to us that \textit{gravity-induced oscillations} provide a better way to measure the influence of gravity on quantum systems, as they do for measuring the influence of gravity on classical systems (i.e., Cavendish's experiment).

Finally, we note that the Hamiltonian (\ref{qubit}) applies to a pair of gravcats in the geometry of Fig. 1. Other geometries will lead to different Hamiltonians, characterized by a different Rabi oscillation frequency $\mho $. Still, $\mho $ will be given by an expression of the form $C Gm^2/D$, where $m$ is the mass of the particles, $D$ the achievable size of a cat state, and $C$ of constant of order unity. In any experimental implementation of the two-gravcat system, we must select the geometry that maximizes $C$, given fixed values for $m$ and $D$.

\section{Controlling the two-level approximation}

In this section, we present a more detailed analysis of the model in Sec. 2. We show how the two-level approximation emerges from standard perturbation theory, and we identify more general expressions for the coefficients through the semiclassical approximation for a general double-well potential. The calculations in this section also serve as templates for analogous calculations in set-ups with different geometry.

\subsection{The interaction matrix}

 We consider the model of Sec. 2  without the simplifying approximation (\ref{xapp}).
We denote  $\psi_0(x):= \langle x|+\rangle$; hence, $\psi_0(-x):= \langle x|-\rangle$. By assumption $\psi_0(x)$ is localized in the right-well, and it decays rapidly in the entire central region and the left well.

The energy split $\omega$ between $E_g$ and $E_e$ is standardly evaluated in the WKB approximation,
\bey
\omega = \Omega L \sqrt{\frac{m \Omega}{\pi}} \exp \left[ \int_0^{\frac{L}{2}}dx \left( \Omega\sqrt{\frac{m}{2U(x)}}-\frac{1}{\frac{L}{2}-x} - 2\sqrt{2mU(x)} \right) \right], \label{omexp}
\eey
where $\Omega$ is the frequency of small oscillations around $x = \frac{L}{2}$, and we have chosen the zero of energy so that $U(\pm \frac{L}{2}) = 0$.

The frequency $\Omega$ defines the order of magnitude for the energy of higher energy levels. Eq. (\ref{omexp}) implies that the ratio $\omega/\Omega$ is exponentially suppressed. This implies that we can treat the gravitational interaction using perturbation theory for almost
degenerate eigenvalues when working on the subspace  $W$ generated by the vectors $|e,e\rangle$, $|g, g\rangle$, $|e, g\rangle$, and $|g, e \rangle$. The contribution from higher eigenstates of the unperturbed system is of the order of $(\omega/\Omega)^2$. Hence, to first order in $G$, it suffices to work on $W$ and to diagonalize the matrix

\bey
V:= \left( \begin{array}{c c c c}
 \omega + V_{ee,ee} & V_{gg, ee} & V_{eg, gg} & V_{ge, gg} \\
V_{ee,gg} &  -\omega + V_{gg, gg} & V_{eg, gg} & V_{ge,gg} \\
 V_{ee,eg} & V_{gg,eg} & V_{eg, eg} & V_{ge,eg} \\
 V_{ee,ge} & V_{gg,ge} & V_{eg,ge} &  V_{ge,ge} \end{array}\right).
\eey

We find that
\bey
\hspace{-1.5cm}  V:= -\alpha (\gamma_+ - \gamma_1) \hat{I} -\alpha \left( \begin{array}{c c c c}
\frac{\omega}{\alpha} +2 (\gamma_1 + \gamma_0) & \gamma_-& 0  & 0 \\
\gamma_-& -\frac{\omega}{\alpha} +2(\gamma_1 - \gamma_0) &0&0 \\
0  &  0 & 0 & \gamma_- \\
0 & 0 &\gamma_- & 0   \end{array}\right), \label{vv2}
\eey
where
\bey
\gamma_{\pm} &=& \frac{1}{2} \int dx_1 dx_2 \frac{\psi_0^2(x_1)\psi_0^2(x_2) \pm \psi_0^2(x_1)\psi_0^2(- x_2) }{\sqrt{d^2+ (x_1-x_2)^2}}, \label{ga+}\\
\gamma_0 &=& \int dx_1 dx_2 \frac{\psi_0^2(x_1)\psi_0(x_2) \psi_0(-x_2) }{\sqrt{d^2+ (x_1-x_2)^2}}, \\
\gamma_1 &=&   \int dx_1 dx_2 \frac{\psi_0(x_1)\psi_0(-x_1)\psi_0(x_2) \psi_0(-x_2) }{\sqrt{d^2+ (x_1-x_2)^2}} \label{ga1}.
\eey

Eq. (\ref{hh}) is obtained from Eq. (\ref{vv2}) in the regime where $\gamma_0$ and $\gamma_1$ are much smaller than $\gamma_- $ and much smaller than $\frac{\omega}{\alpha}$. Eq. (\ref{vv2}) allows us to control the approximations effected in Sec. 2, once the wave function $\psi_0(x)$ is specified.

\subsection{Approximate evaluation of the interaction matrix}

A good approximation is obtained by assuming that $\psi_0$ is close to the ground state of a harmonic oscillator with frequency $\Omega$ around $x = \frac{L}{2}$ \cite{Garg}. Then, in the classically allowed region,
\bey
\psi_0(x) = \left(\frac{m \Omega}{\pi}\right)^{1/4}e^{-\frac{m\Omega}{2}(x - \frac{1}{2}L)^2}. \label{psi0osc}
\eey
The classically forbidden region corresponds to $ -a<x< a$, where $a$ is defined by $U(a) = \frac{1}{2} \Omega$. In this region, $\psi_0(x)$ is approximated by the WKB expression
\bey
\psi_0(x) = \sqrt{\frac{m\omega}{2}} \left(2m[U(x') - \frac{1}{2}\Omega]\right)^{-1/4}  e^{\int_0^{x} dx' \sqrt{2m[U(x') - \frac{1}{2}\Omega}]}.
\eey
We also take $\psi_0$ to be negligible for $ x < - a$.

For $\sqrt{m\Omega} L >> 1$, we can evaluate $\gamma_+$ and $\gamma_-$ by substituting Eq. (\ref{psi0osc}) for $\psi_0$ everywhere, because the contribution from the classically forbidden area is negligible. We find,
\bey
\gamma_+ + \gamma_- = \sqrt{\frac{m\Omega }{\pi}} e^{\frac{m\Omega d^2}{8}} K_0(\frac{m\Omega d^2}{8})\\
\gamma_+ - \gamma_- = \frac{2}{\sqrt{d^2+L^2}} + O(\frac{1}{\sqrt{m\Omega} L}).
\eey

In this regime, the frequency  $\mho$ of the Rabi oscillations, Eq. (\ref{u}), is given by the more accurate expression
\bey
\mho = \alpha \left[ \frac{1}{2}\sqrt{\frac{m\Omega }{\pi}} e^{\frac{m\Omega d^2}{8}} K_0(\frac{m\Omega d^2}{8}) - \frac{1}{d'}\right],
\eey
which is valid also in the regime where $d << L$.

To evaluate $\gamma_0$ and $\gamma_1$, we note that for $\sqrt{m\Omega} L >> 1$, the dominant contribution to $\psi_0(x) \psi_0(-x)$ comes from the term in the forbidden region, where
\bey
\psi_0(x) \psi_0(-x) = \frac{\omega}{2} \sqrt{\frac{m}{2[U(x) - \frac{1}{2} \Omega]}}.
\eey
In this regime, the turning point $a$ is very close to $\frac{L}{2}$, so we can approximate
\bey
\gamma_1 &=& \frac{m\omega^2}{8} \int_{-L/2}^{L/2}dx_1 \int_{-L/2}^{L/2}dx_2 \frac{1}{\sqrt{U(x_1) U(x_2)} \sqrt{d^2+ (x_1-x_2)^2 }} \\
\gamma_0 &=& \frac{\omega \sqrt{m}}{2\sqrt{2}}  \int_{-L/2}^{L/2}dx  \frac{1}{\sqrt{U(x)} \sqrt{d^2+ (x_1-x_2)^2 }  }.
\eey

\subsection{Two-dimensional potential}

In the models above, we assumed that the two particles move along one dimension, along parallel axes of fixed distance $d$. This is obviously an idealization. In a realistic set-up, the particles are confined by a potential. This is possible for a two-dimensional configuration with a potential
\bey
U(x, y) = U_1(x) + U_2(y). \label{extpot}
\eey
The potential $U_1(x)$ is the one we considered in the previous sections. The potential $U_2(y)$ has two sharp minima at $y = \pm \frac{d}{2}$, but its height is much larger than the height of $U_1$. This means that  transitions through tunneling between the two minima is negligible compared to tunneling transitions in $U_1$.

Furthermore, we assume that the associated localization length $\xi$ in the $y$ direction is much smaller than $d$. If the initial state consists of one particle localized at $y = \frac{d}{2}$ and a second particle localized at $y = - \frac{d}{2}$, then the results of the previous sections arise as leading order terms with corrections of the order of $\frac{\xi}{d}$.

In this set-up, the state (\ref{grrr}) is not the true ground state of the system, but a very long-lived metastable state. The true ground state contains an amplitude in which both particles can be in the same well. In this case, they may interact through channels other than  Newtonian gravity, e.g., contact interactions, van der Waals forces  and so on.

\section{QFT description of the gravcats}

\subsection{The Hamiltonian}
Consider a non-relativistic scalar field in two dimensions corresponding to particles of mass $m$. The single particle Hilbert space   ${\cal H} = L^2({\pmb R}^2, d^2p)$ consists of square integrable momentum wave-functions. The field operators are defined on the associated Fock space ${\cal F}({\cal H})$. The field operators are simply the Fourier transforms of the creation and annihilation operators on the Fock space,
\bey
\hat{\psi}({\pmb x}) = \int \frac{d^2p}{2\pi} \hat{a}({\pmb p})e^{i {\pmb p} \cdot {\pmb x}}, \hspace{1cm}\hat{\psi}^{\dagger}({\pmb x}) = \int \frac{d^2p}{2\pi} \hat{a}^{\dagger}({\pmb p})e^{-i {\pmb p} \cdot {\pmb x}}.
\eey
The field Hamiltonian for two-particle interactions is
\bey
\hat{H} &=& \int d^2x \left[-\frac{1}{2m} \hat{\psi}^{\dagger}({\pmb x}) \nabla^2 \hat{\psi}({\pmb x}) + \hat{\psi}^{\dagger}({\pmb x})U({\pmb x})\hat{\psi}({\pmb x})\right]
\nonumber \\
&+& \frac{1}{2} \int d^2x d^2x' V({\pmb x}- {\pmb x'}) \hat{\psi}^{\dagger}({\pmb x})\hat{\psi}({\pmb x})\hat{\psi}^{\dagger}({\pmb x'})\hat{\psi}({\pmb x'}),
\eey
where $U$ is an external potential of the form (\ref{extpot}). The self-interaction potential
\bey
V({\pmb x}) = V_N({\pmb x}) + V_s({\pmb x})
\eey
 is the sum of the  Newtonian interaction
\bey
V_N({\pmb x}) = - \frac{\alpha}{|{\pmb x}|},
\eey
and of a short-range potential $V_s({\pmb x})$. In what follows, we will assume that the range $r_0$ of $V_s$ is much smaller than all physically relevant length-scales so that we can effectively treat $V_s$ as a delta function
\bey
V_s({\pmb x}) = \frac{1}{2} g \delta^2({\pmb x}).
\eey
The coupling constant $g$ is proportional to the scattering length of a pair of bosons.

Let us assume that the potential $U_2$ along the $y$ direction restrict the motion of the particles only along the axes $y = \pm \frac{d}{2}$---in the sense explained in Sec. 3.3 . Then, we can split the Hilbert space ${\cal H}_1$ as ${\cal H}_1 \oplus {\cal H}_2$, where ${\cal H}_{1(2)}$ contains states with support only on positive (negative) values of $y$. Since
\bey
{\cal F}({\cal H}_1 \oplus {\cal H}_2) = {\cal F}({\cal H}_1) \otimes {\cal F}({\cal H}_2),
\eey
the field behaves as a bipartite system. We express the  fields on ${\cal F}({\cal H}_1)$ as $\hat{\psi}_1({\pmb x})$ and $\hat{\psi}^{\dagger}_1({\pmb x})$ and the fields on ${\cal F}({\cal H}_2)$ as $\hat{\psi}_2({\pmb x})$ and $\hat{\psi}^{\dagger}_2({\pmb x})$. Note that the associated creation and annihilation operators are not labeled by ${\pmb p}$ because the momentum does not define a generalized basis on ${\cal H}_{i}$.

We reduce each subsystem to motion along the $x$ axis, ignoring the small fluctuations around $y = \pm\frac{d}{2}$ allowed by the potential $U_2$.
Then, the system is equivalent to a pair of one-dimensional non-relativistic fields, with Hamiltonian
\bey
\hat{H} = (\hat{h}_1 + \hat{\upsilon}_1) \otimes \hat{I} + \hat{I} \otimes (\hat{h_2} + \hat{\upsilon}_2) + \hat{V}, \label{Htot}
\eey
where
\bey
\hat{h}_i = \int dx \left[ \hat{\psi}_i^{\dagger}\left(-\frac{1}{2m} \partial_x^2 + U_1\right) \hat{\psi}_i  + \frac{1}{2} g :(\hat{\psi}_i^{\dagger}\hat{\psi}_i)^2:\right] \label{hii}
\eey
is the  Hamiltonian for a Bose-Einstein gas in one dimension,
\bey
\hat{\upsilon}_i = - \frac{1}{2} \alpha \int dx dx' \frac{ :\hat{\psi}_i^{\dagger}( x)\hat{\psi}_i(x)\hat{\psi}_i^{\dagger}(x')\hat{\psi}_i(x'):}{|x-x'|}, \label{vii}
\eey
is the gravitational self-interaction term for the Bose gas,
and
\bey
\hat{V}_N = - \alpha \int dx_1 dx_2  \frac{ \hat{\psi}_1^{\dagger}( x_1)\hat{\psi}_1(x_1)\hat{\psi}_2^{\dagger}(x_2)\hat{\psi}_2(x_2)}{\sqrt{d^2+(x_1-x_2)^2}}. \label{intg}
\eey
is the gravitational interaction between the two subsystems.

We have renormalized $\hat{h}_i$ and $\hat{V}_i$ by requiring that its expectation value on the vacuum vanishes: this is straightforwardly achieved by taking all terms to be normal-ordered.

The Hamiltonian (\ref{Htot}) commutes with both particle-number operators $\hat{N}_i = \int dx \hat{\psi}^{\dagger}_i\hat{\psi}_i$. In non-relativistic quantum field theory, particle numbers define superselection sectors. This implies that the Hilbert space for one field splits into subspaces ${\cal H}_N$ indexed by particle numbers $N$. Accordingly, the   Hilbert space of our bipartite system splits into subspaces   ${\cal H}_{N_1, N_2} = {\cal H}_{N_1} \otimes {\cal H}_{N_2}$, indexed by the particle numbers $N_1, N_2$. There are no superpositions between different subspaces. Hence, in all physical situations, we can consider $N_1$ and $N_2$ to be fixed parameters.

\subsection{The Gravitational Gross-Pitaevski equation}

The Hamiltonian $\hat{h}+ \hat{\upsilon}$   describes the gravitational self-interaction of an one-dimensional Bose-Einstein gas. For $N >> 1$, this system is well described by the   Hartree (mean field) approximation.   We consider $N$-particle states of the form $|\Psi(\phi)\rangle = |\phi\rangle \otimes |\phi\rangle \otimes \ldots \otimes |\phi\rangle$, where $|\phi\rangle$ corresponds to an one particle wave-function $\phi(x)$. Stationary states are obtained by the variation of the energy functional $E[\phi]  = \langle \Psi(\phi)|\hat{h}|\Psi(\phi)\rangle$.
The result is the Hartree-type equation
\bey
-\frac{1}{2m} \partial_x^2 \phi(x) + U_1 \phi(x) &+& g(N-1) |\phi(x)|^2 \phi(x) \nonumber \\
&-& \alpha (N-1) \phi(x) \int dx' \frac{|\phi(x')|^2}{|x-x'|} = \mu \phi(x), \label{GPN}
\eey
where $\phi$ is normalized as $\int dx |\phi(x)|^2 = 1$, and $\mu$ is the chemical potential.

Eq. (\ref{GPN}) is the \textit{Gravitational Gross-Pitaveskii} (GGP) \textit{Equation}. It reduces to the Gross-Pitaevskii equation if the gravitational self-interaction is much smaller than inter-particle interaction (set $\alpha \simeq 0)$, and it reduces to the so-called Newton-Schr\"odinger  equation (NSE) in the opposite regime ($g \simeq 0$)\footnote{As explained in Ref. \cite{AHNSE}, the NSE makes sense from the viewpoint of GR and QFT only under the Hartree  approximation that is valid for $N >> 1$. In contrast, the popular NSE for $N = 1$ or small is an alternative quantum theory not derivable from GR+QFT.}.

%By continuity, we expect that for weak couplings $g$ and $\alpha$, Eq. (\ref{GPN}) admits a ground state solution $\phi_g(x)$ that is parity symmetric and an excited solution $\phi_e(x)$ that is parity-antisymmetric, with a small difference $\omega$ in the associated energies. We then define

\subsection{The two-mode approximation}
%The results of Sec. 2 apply for $N_1= N_2 = 1$. They can be generalized to the regime $N_1,  N_2 >> 1$, relevant to a Bose-Einstein condensate, as follows.
We will follow the two-mode approximation \cite{MCWW, OKLHCA} that is commonly employed for BECs in a double-well potential. To this end, we focus on the Gross-Pitaevskii equation
\bey
-\frac{1}{2m} \partial_x^2 \phi(x) + U_1 \phi(x) + g(N-1) |\phi(x)|^2 \phi(x)  = \mu \phi(x), \label{GP0}
\eey
and  assume that  $g(N-1) $ is sufficiently small  so that the ground and first excited state of (\ref{GP0}) have similar characteristics to the solution of Schr\"odinger's equation, employed in Sec. 2. In particular, we assume that the real-valued ground state wave function $\phi_0(x)$ is parity-even, that the real-valued wave function of the first-excited state is parity odd, and that the difference in the  chemical potentials (eigenvalues) for these solutions is small. This implies that the function
$\psi_0 = \frac{1}{\sqrt{2}}(\phi_0+\phi_1)$  is well localized in the right well.
%For $g = 0$, $\psi_0$ coincides with the function of the same name, described in Sec. 3.

The two mode approximation consists in expressing the quantum field $\hat{\psi}(x)$ as
\bey
\hat{\psi}(x) = \hat{a}_0 \phi_0(x) + \hat{a}_1 \phi_1(x). \label{twomode}
\eey
Then, the Hamiltonian is expressed solely in terms of the four bosonic operators $\hat{a}_0, \hat{a}_1, \hat{a}^{\dagger}_0$ and $\hat{a}^{\dagger}_1$, subject to the constraint that the total particle number is equal to $N$: $\hat{a}^{\dagger}_0 \hat{a}_0 + \hat{a}^{\dagger}_1 \hat{a}_1 = N \hat{I}$.  This Hilbert space defines   a representation of SU(2) with $j = \frac{N}{2}$.The angular momentum generators related to the bosonic operators through the Schwinger correspondence
\bey
\hat{S}_x = \hat{a}_1^{\dagger}\hat{a}_0 + \hat{a}^{\dagger}_0 \hat{a}_1, \\
\hat{S}_y = i( \hat{a}_1^{\dagger}\hat{a}_0 - \hat{a}^{\dagger}_0 \hat{a}_1 ) \\
\hat{S}_z = \hat{a}_1^{\dagger}\hat{a}_1 - \hat{a}^{\dagger}_0 \hat{a}_0.
\eey
Hence, the Hilbert space of the system is spanned by the $N+ 1$  eigenstates $|m\rangle$ of $\hat{S}_z$, where $m = -\frac{N}{2}, -\frac{N}{2}+1, \ldots, \frac{N}{2} - 1, \frac{N}{2}$.

Substituting into Eq. (\ref{hii}), we obtain
\bey
\hspace{-2cm} \hat{h} =  \omega \hat{S}_z + g \left[ \frac{1}{4}N^2 (\delta_0 + \delta_1) \hat{I} - \frac{1}{2}N(\delta_0+3\delta_1) \hat{I} + \frac{1}{2}\delta_1 \hat{S}_z^2 +\frac{1}{4}   (\delta_0 - \delta_1)  \hat{S}_x^2\right],
\eey
where $\omega = \mu_2 - \mu_1$ is the tunneling oscillation frequency,
\bey
\delta_0 = \int dx \psi_0^4(x), \hspace{1cm} \delta_1 = \int dx \psi_0^2(x) \psi_0^2(-x).
\eey

The gravitational self-interaction term (\ref{vii})  is
\bey
\hat{\upsilon} = -\frac{1}{2} \alpha \left[  N^2 \beta_+ \hat{I} - 2(N+1) \beta_0 \hat{S}_z +   \beta_1 \hat{S}_z^2 +  \beta_-  \hat{S}_x^2 \right],
\eey
where
\bey
\beta_{\pm}& =& \frac{1}{2} \int dx dx' \frac{\psi_0^2(x) \psi_0^2(x') \pm \psi^2_0(x)\psi_0^2(-x')  }{|x-x'|},\\
\beta_0 &=& \int dx dx' \frac{\psi_0^2(x) \psi_0(x')\psi_0(-x')  }{|x-x'|}\\
\beta_1 &=& \int dx dx' \frac{\psi_0(x) \psi_0(-x) \psi_0(x')\psi_0(-x')  }{|x-x'|}.
\eey

We ignore terms proportional to unity to write

\bey
 \hat{h} + \hat{\upsilon} =  \tilde{\omega}  \hat{S}_z + \frac{1}{2}(g\delta_1 - \alpha \beta_1)  \hat{S}_z^2 + \frac{1}{2} (g \delta_- - \alpha \beta_-) \hat{S}_x^2,
\eey
where  $\delta_- = \frac{1}{2} (\delta_0 - \delta_1)$  and   $\tilde{\omega} = \omega + \alpha (N+1) \beta_0$. The term
\bey
\Delta \omega:= \alpha (N+1) \beta_0,
\eey
is the Lamb shift of the tunneling frequency due to the Newtonian self-interaction.

Note that for $N = 1$, $\hat{S}_z^2 = \hat{S}_x^2 = \frac{1}{4}\hat{I}$, hence, all terms except for $\hat{S}_z$ can be absorbed in a redefinition of energy scale. The self-interaction term in the GP equation also vanishes at $N = 1$.   Hence, we recover the results of Sec. 2.

In an early discussion of superpositions of the gravitational field of the BEC \cite{Peres}, the BEC in a double well potential is treated as a two-level system, where the two states correspond to either all particles in the right well or all particles in the left well. This is a subspace of the Hilbert space considered here, spanned by the vectors with quantum number $m = - \frac{N}{2}$ and $m = \frac{N}{2}$.

Experiments with BECs were proposed in Ref. \cite{FP}for testing gravity-induced quantum state reduction. Howl, Penrose and Fuentes \cite{HPF}  calculated the self-energy of the difference between spherical and spheroidal BEC mass distributions, and compare  the corresponding rate of state reduction to the environmental  decoherence rate  in BEC experiments. They  provide estimates for the values of experimental parameters, such as temperature and scattering length, that would be required to test  gravitational quantum state reduction.

Next, we consider a pair of BECs corresponding to the subsystems $1$ and $2$ of Sec. 4.1. For simplicity, we consider equal number of particles in the
two subsystems ($N_1 = N_2 = N$). The
  interaction term (\ref{intg})   is
\bey
\hspace{-1.5cm} \hat{V}_N =  -\alpha \left[ \frac{N^2}{4} \gamma_+ \hat{I} -N \gamma_0 (\hat{S}_z\otimes \hat{I} + \hat{I}\otimes \hat{S}_z ) +    \gamma_1 \hat{S}_z \otimes \hat{S}_z + \gamma_- \hat{S}_x \otimes \hat{S}_x                            \right],
\eey
where $\gamma_{\pm}, \gamma_0$, and $\gamma_1$ are given by Eqs. (\ref{ga+}---\ref{ga1}).

Dropping the constant terms, we write the  Hamiltonian for the bipartite system as
\bey
\hat{H} = \hat{H}_0\otimes \hat{I} + \hat{I} \otimes \hat{H}_0 + \hat{H}_I,
\eey
where
\bey
\hat{H}_0 &=& \bar{\omega}  \hat{S}_z+ \frac{1}{2} (g\delta_1 - \alpha \beta_1)   \hat{S}_z^2 + \frac{1}{2} (g \delta_- - \alpha \beta_-) \hat{S}_x^2,\\
\hat{H}_I &=& - \alpha \gamma_1 \hat{S}_z \otimes \hat{S}_z  - \alpha  \gamma_- \hat{S}_x \otimes \hat{S}_x.
\eey
The frequency $\bar{\omega}$ includes a contribution $\Delta \omega_I = \alpha N \gamma_0$ from the Newtonian interaction between the two subsystems , $\bar{\omega} = \tilde{\omega} +
\Delta \omega_I $.

The equations above apply to
\begin{itemize}
\item all $N \geq 1 $, if $g = 0$,
\item $N >> 1$, if $g \neq 0$.
\end{itemize}

As discussed in Sec. 3, the physically relevant regime corresponds to $\sqrt{m\Omega} L >> 1$. For sufficiently small value of $gN$, so that the solution $\psi_0$ is well concentrated in the right well
$\delta_1 << \delta _0$, $\beta_1 << \beta_-$, and $ \gamma_1 << \gamma_-$. Then, we can write
\bey
\hat{H}_0 = \bar{\omega}  \hat{S}_z +\frac{1}{2} \kappa \hat{S}_x^2, \label{H02} \\
\hat{H}_I = -  \mho \hat{S}_x \otimes \hat{S}_x \label{HI2}
\eey
where we wrote $ \kappa =  (g \delta_- - \alpha \beta_-)$ and
$\mho =   \alpha  \gamma_-$.

The expressions above generalize Eq. (\ref{qubit}) for a pair of $N$-particle systems trapped in a double-well potential.

\subsection{Physical effects}
The field operator (\ref{twomode}) can be written as
\bey
\hat{\psi}(x) = \frac{1}{\sqrt{2}}(\hat{a}_0+\hat{a}_1) \psi_0(x) + \frac{1}{\sqrt{2}}(\hat{a}_0 - \hat{a}_1) \psi_0(x).
\eey
The operator $\hat{a}_R = \frac{1}{\sqrt{2}}(\hat{a}_0+\hat{a}_1)$ can be interpreted as an annihilation operator for the particle in the right well, and the operator $\hat{a}_L = \frac{1}{\sqrt{2}}(\hat{a}_0 - \hat{a}_1)$ as an annihilation operator for the particle in the left well. We define the associated particle-number operators $\hat{N}_L = \hat{a}^{\dagger}_L \hat{a}_L$ and $\hat{N}_R = \hat{a}^{\dagger}_R \hat{a}_R$. Then, $\hat{N}_L + \hat{N}_R = N \hat{I}$ and $\hat{N}_R - \hat{N}_L = 2 \hat{S}_x$. Hence, the operator $\hat{S}_x$ determines the number of particles in the two wells.

A quantum rotor with $j >> 1$ is well approximated by its classical analogue, i.e., a Hamiltonian system on the two-sphere $S^2$. Hence, for $N>>1$, we can understand the BEC dynamics in the double-well potential by using the classical equations of motion. To this end, we substitute all spin operators with their associated classical functions. It is convenient to use coordinates $\xi \in [-1,1)$, and $\varphi \in[0, 2\pi)$ on the sphere, such that
\bey
S_x = N \xi \\
S_y = N\sqrt{1-\xi^2} \sin \varphi \\
S_z = - N \sqrt{1-\xi^2} \cos \varphi.
\eey
The functions above satisfy the SU(2) algebra with respect to the Poisson bracket
\bey
\{ \varphi, \xi \} = N^{-1}. \label{poiss}
\eey
The symplectic form on the two sphere is $\omega = N \sin \theta d \theta \wedge d \varphi$, in terms of the usual coordinates $(\theta, \varphi)$ of the two-sphere. The coordinates $(\xi, \varphi)$ coincide with $(\cos \theta, \varphi)$ after  relabeling the axes.

The classical Hamiltonian that corresponds to (\ref{H02}) is
\bey
H_0(\xi, \varphi) = N \bar{\omega} \left( -\sqrt{1-\xi^2} \cos \varphi + \frac{1}{2}b \xi^2\right).
\eey
where $b = N \kappa/\bar{\omega}$.

For $b > 0$, the absolute minimum of $H_0$  is at $ \xi = 0,  \varphi = 0$. The Hamiltonian for small oscillations around that minimum is
\bey
H_0(\xi, \varphi) = N\bar{\omega} \left[ \frac{1}{2} (1 + b) \xi^2 + \frac{1}{2} \varphi^2\right], \label{h0cl}
\eey
where we dropped the constant term $-N \bar{\omega}$. Given Eq. (\ref{poiss}), Eq.  (\ref{h0cl}) describes a harmonic oscillator with frequency $\Omega_0 =  \bar{\omega} \sqrt{1+b}$.

The total Hamiltonian for the system of two identical  rotors interacting through Eq. (\ref{HI2}) is the function
\bey
\hspace{-2.5cm} H(\xi_1, \xi_2, \varphi_2, \varphi_2) = N \bar{\omega} \left[-\sqrt{1-\xi_1^2} \cos \varphi_1 -\sqrt{1-\xi_2^2} \cos \varphi_2+ \frac{1}{2}b (\xi_1^2+\xi_2^2) -  c \xi_1 \xi_2\right] \label{Ham2O}
\eey
defined on the manifold $S^2 \times S^2$. We wrote $ c = N\mho/\bar{\omega}$.

For small oscillations near the energy minimum at $\xi_1 = \xi_2 = \varphi_1 = \varphi_2 = 0$,
\bey
H(\xi_1, \xi_2, \varphi_1, \varphi_2)  = \frac{1}{2} N\bar{\omega} \left[(1+b)(\xi_1^2 + \xi_2^2) + \varphi_1^2 +\varphi_2^2 - 2c \xi_1 \xi_2 \right].
\eey
 This Hamiltonian leads to a coupling between the two sets of degrees of freedom, with characteristic frequencies $\Omega_{\pm} =  \bar{\omega} \sqrt{1 + b \pm c}$. This system of coupled rotors is characterized by periodic energy transfer between the two subsystems. The associated frequency $\delta$ equals  $\Omega_+ - \Omega_-$, and it    increases with $c$, as can be seen from Fig. 2.

For $c << 1$ or $c << b$, $ \delta = 2 N \mho/\sqrt{1+b}$. Systems of two interacting rotors are known to have chaotic behavior for intermediate values of their angular momenta \cite{FePe}. We have not found such behavior in the regimes we have considered. A thorough analysis of the model's parameter space is necessary, in order to identify or rule out a quantum chaos regime.

In the BEC system, the frequency of  gravity induced oscillations increases with the number of particles $N$ in the condensate. BECs are less efficient in demonstrating such oscillations than individual heavy particles with the same mass: the Rabi frequency of the BEC increases linearly with the total mass $M = Nm$, while it increases quadratically for a single particle.
For a BEC of Rb atoms,  size of the cat state of the order of $100 nm$, and $\delta^{-1}$ of the order of one minute, we need a very large number $N \sim 10^{16}$ of atoms.

There are other ways to see the effect of the gravitational interaction in a BEC. Consider a set-up, in which the subsystem 1 is prepared in the ground state, and the subsystem 2 is well excited. From elementary classical mechanics of coupled oscillations, we find that the subsystem 1 receives energy with rate
\bey
\left|\frac{dh_1}{dt}\right| \sim \bar{\omega} N^2 \mho ,
\eey
for time much larger than $\omega^{-1}$ but much smaller than $\delta^{-1}$. Hence, the number  $n_1 = h_1/\bar{\omega}$ of excited atoms in subsystem $1$ is of the order of $N^2 \mho t$, i.e., $n_1$ increases quadratically with $N$. (This corresponds to the very early time increase of the functions in Fig. 2.) Hence, early time transfer of energy between the two subsystems is proportional to the total mass of each subsystem both for individual heavy particles and for BECs. For a BEC of Rb atoms,  size of the cat state of the order of $100 nm$, and $t \sim 5s$, a macroscopically large value of $n_1 \sim 10^3$ requires BECs with $N \sim 10^{10}$ atoms.

In general, the measurement of any dynamical correlation between the two subsystems will be evidence for the gravitational interaction. Such correlations can be predicted from the dynamics generated by Eqs. (\ref{H02}---\ref{HI2}).

   \begin{figure}
    \centering

 \includegraphics[width=0.6\textwidth]{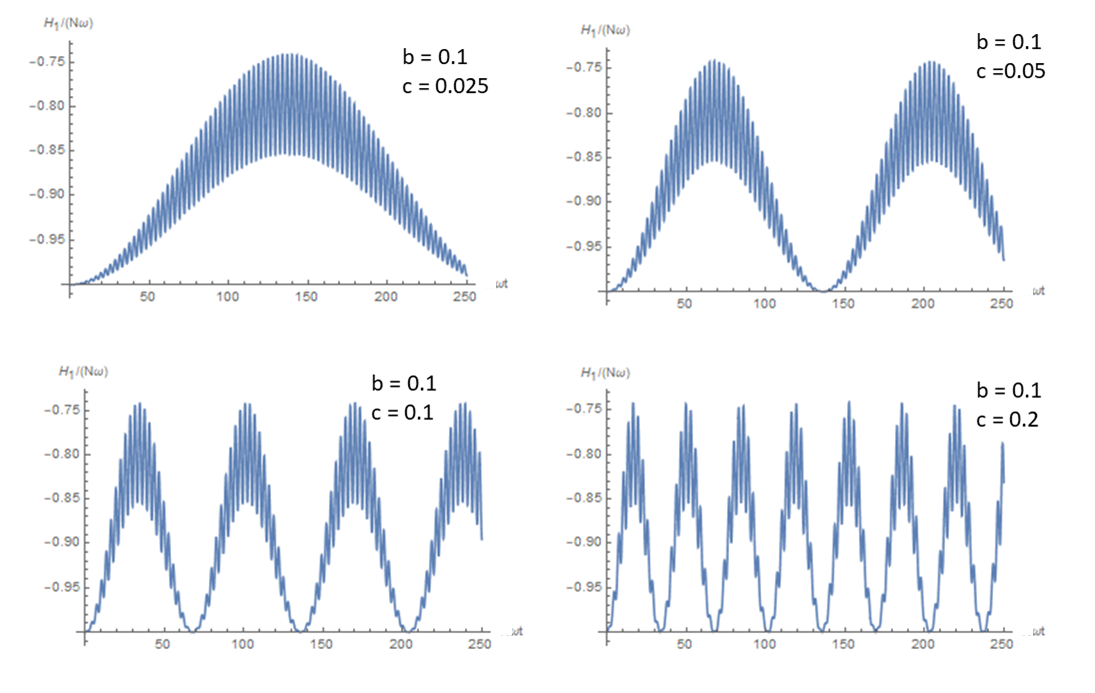}

    \caption{Energy transfer between the two rotors, derived from Hamilton's equations for the full Hamiltonian (\ref{Ham2O}). We plot the energy $h_1 /( N\bar{\omega}) = \frac{1}{2} (1 + b) \xi^2 + \frac{1}{2} \varphi^2$ of the first rotor as a function of dimensionless time $\bar{\omega}t$. We assume that the first rotor is initially in its ground state and that the second is excited. We choose initial conditions are $\varphi_1 = 0, \xi_1 = 0, \varphi_2 = 0, \xi_2 = 0.5$, and we consider different values of $b$ and $c$. }
\end{figure}

\section{Discussion: Physical interpretation}

Before ending we want to discuss three aspects related to the present work: 1) What physics, relevant to the quantum nature of gravity especially, can experiments following the set-ups described here, expound? For this we need a preamble making precise the definition of `quantum gravity' and the meaning of the `quantum nature' of gravity.  2) The relevance of our findings to alternative quantum theories. 3) Fundamental theoretical issues pertaining to this class of phenomena.

\subsection{What physics could experiments of this type demonstrate}

On the theory side, this line of investigation involves three main threads: gravity, quantum and information. Quantum information issues of interest to us include decoherence and entanglement. Interestingly when it comes to bringing the other two familiar subjects together, that of gravity and quantum,  some clarification proves necessary because the term `quantum gravity' is used  loosely in claims of what tabletop experiments \cite{Carney} can deliver. Because of this we need a preamble, to make precise the physical meanings of terminology used, and to define clearly the contexts and scopes to avoid possible confusion over stated goals.

\subsubsection{Preamble:  Define `Quantum Gravity' more precisely}

Let us agree that when only `gravity' is mentioned it is taken to be classical gravity.  Classical gravity is described with very high accuracy by Einstein's general relativity theory.  One can distinguish two domains, weak field, such as experiments on earth or in the solar system would fall under, from strong field, such as processes near black holes or neutron stars,  depending on their masses and the proximity of measured events, and in the early universe.

Classical \textit{ `perturbative gravity'}  refers to small perturbations off of a classical gravitational background spacetime.  In the earth's environments, the background spacetime is Minkowski space. While the background spacetime could be strongly curved, perturbative treatments can only consider small amplitude deviations which are weak by proportion. Gravitational waves  are usually treated as perturbations  whose wavelengths can span from the very long of astrophysical or cosmological scales to the very short. Note the crucial difference between \textit{perturbations} and \textit{fluctuations}, the former being a deterministic variable, referring to the small amplitude deviations from the background spacetime, while the latter is a stochastic variable,  referring to the noise. Fluctuations in the gravitons constitute a noise of gravitational origin and of a quantum nature. Such effects at low energies  are in principle detectable from gravitational wave observatories \cite{PWZ}.  Fluctuations of quantum matter field can also induce metric fluctuations. They are governed by the Einstein-Langevin equation of semiclassical stochastic gravity theory \cite{HuVer20}. At the Planck scale they make up the spacetime foams.

 \textit{`Quantum Gravity'  proper}  refers to theories of the basic constituents of spacetime at the near-Planckian scale, such as string theory, loop quantum gravity, spin network, causal dynamical triangulation, asymptotic safety, causal sets, group field theory, spacetime foams, etc \cite{Oriti}. Because of the huge  energy scale discrepancy between the Planck scale and the scale of a laboratory or a satellite laboratory, many such experimental proposals to test Planck scale quantum gravity need to rely on \textit{indirect} implications to high-energy particle phenomenology \cite{QGPhen,Sabine} or, at a lower energy range, analog gravity experiments \cite{AnalogG}. Of the latter kind, many atomic-molecular-fluid, condensed matter-BEC or electro-optical-mechanical experiments can indeed skillfully use analogs to seek indirect implications of quantum gravity. But in terms of \textit{direct} observations, or drawing direct implications, tabletop experiments can only probe weak-field perturbative gravity, nevermind their quantum gravity labels.

 \textit{Gravitons} are the quantized propagating degrees of freedom of weak perturbations off of  a background metric,   such as the Minkowski spacetime for experiments in the Earth's environment.
Graviton as a \textit{spin 2 particle} refers to the high frequency components of weak gravitational  perturbations under certain averages (like the Brill-Hartle-Isaacson average), or in the ray representation under the eikonal approximation.

% As such, beware of the shortcomings when relying on gravitons  (lacking phase information)  to address  fully the quantum information issues of perturbative quantum gravity:  quantum coherence,  correlations and entanglement between gravitons need be considered as a whole.

 Gravitons can exist at today's very low energy and thus have little to do with the basic constituents of spacetime at the Planck energy. One does not need any new deeper level theory for their description. Einstein's general relativity theory plus second quantization on weak linear perturbations will do, much as photons in QED based on Maxwell's electromagnetism. Note that any linearized degree of freedom in classical systems can be quantized, irrespective of whether it is fundamental or collective. The latter is in abundance in condensed matter physics (e.g., phonons, rotons, plasmons, and many other entities with -on endings).
Seeing the quantum nature of the gravitational field at today's low energy, such as proving the existence of gravitons \cite{Dyson,PWZ},  is certainly of fundamental value, but it still has no bearing on  quantum gravity proper as defined above. Gravitons are the quantized excitation modes of spacetime, not the basic building blocks of spacetime. Graviton's existence is predicated on the emergence of spacetime while the building blocks (such as strings, loops, causets, etc) are the progenitors of spacetime structure.

%The term `quantum gravity' attached to experiments or observations is increasingly used by authors from all fields of physics and beyond,  while, its alluring appeal notwithstanding, in actuality what they can say is likely only about the quantum nature of perturbative gravity.

\subsubsection{Pure gauge says nothing about quantum nature}

 A more subtle yet important misconception is attributing quantumness to the pure gauge degrees of freedom (Newton or Coulomb forces), while only the dynamical degrees of freedom (graviton or photon) are the true signifiers of the quantum nature of a theory. Experiments measuring the entanglement between two quantum objects, like the ones proposed in this paper, and in Refs. \cite{Bose17, Vedral17}, albeit through Newtonian gravitational interactions, only expresses the quantum nature of these objects, not of gravity.
This critique is raised in \cite{AnHucr}. There, the authors use the  universally familiar case of electrodynamics to illustrate this point, namely, the  Coulomb term is pure gauge and slaved to matter. Only  the dynamical degrees of freedom, the photons, signify the quantum nature of electromagnetism (EM).  Below is a brief summary of the main points presented there.

In the interaction of electrically charged matter fields $\psi$ with the EM field, the system has a first-class constraint, namely, Gauss law: $\partial_aE^a = 4 \pi \rho$, where $\rho$ is the charge density of the $\psi$ fields. First-class constraints split the dofs of a system into {\em pure-gauge} and {\em true} (i.e., physical)  ones. The former dofs are not observable and they must be removed from  the dynamical description
\cite{constraints}. In EM theory, the longitudinal components of $E^a$ and $A_a$ are pure gauge, and the transverse ones (EM waves) are the true dofs. Only the latter appear in the Hamiltonian
\bey
 H = H_{\psi} + \int d {\bf r} d{\bf r'} \frac{\rho({\bf r}) \rho({\bf r'})}{ |{\bf r} - {\bf r'}|} +H_T +H_{I},
 \eey
%\label{Hamil2}\end{eqnarray}
where $H_{\psi}$, $H_T$ and $H_I$ describe  matter,  EM  waves and   matter-radiation interaction, respectively.

The constraint analysis  generates the Coulomb term %in Eq. (\ref{Hamil2})
through the removal of the pure-gauge dofs.  The Coulomb term depends only on---it is `slaved' to---the matter dofs. Hence, it can exist without the true dofs of the EM field.    For quantum matter, the charge density is represented by an operator. Then, the Coulomb term is assimilated into  the matter Hamiltonian (e.g., as the potential term in the hydrogen atom Hamiltonian).  Suppose we prepare  a bipartite system that consists of two spatially-separated charge distributions. Then, the Coulomb term {\em alone} generates entanglement between the two components, irrespective of whether  the dynamical dofs, the EM-waves, have been quantized or not. Conversely, the generation of entanglement through the Coulomb term says nothing about the quantum nature of the EM field.

The same holds for weak gravity. The linearization of the GR field equations around Minkowski spacetime leads to an action similar to the EM one. The true dofs of the linearized perturbations are their transverse-traceless (TT) components, i.e., gravitational waves (GWs). The quantization of the short-wavelength TT perturbations  gives rise to gravitons.

The Hamiltonian in the weak gravity (WG)-nonrelativistic (NR) limit is
\begin{eqnarray}
 H = H_{mat} -    G \int d{\bf r} d{\bf r'} \; \frac{\mu({\bf r}) \mu({\bf r '})}{|{\bf r} - {\bf r'}|} +H_{TT} + H_{I}
\label{hlin2}  \end{eqnarray}
where $\mu({\bf r})$ is the mass density; $H_{mat}$, $H_{TT}$ and  $H_{I}$ describe matter, GWs and GW-matter  interaction respectively.
The Newtonian interaction term in the Hamiltonian above %in Eq. (\ref{hlin2})
is generated by a pure-gauge component $\phi$ of the perturbations  through the    Poisson equation $\nabla^2 \phi = -4 \pi G \mu$. The latter represents the  scalar constraint of GR in the WG-NR limit \cite{AHNSE}.  The  above arguments for the EM field can be straightforwardly repeated here, to the effect that entanglement generated by the quantized Newtonian term is agnostic  to the nature of the gravitational true dofs, classical or quantum, (i.e., to the existence of gravitons).   For the same reason, a detection of gravcats as in this paper,  in Ref. \cite{AnHu15} or Refs. \cite{Bose17, Vedral17} is also no indication of the quantum nature of gravity.

Formally, the Hilbert space for the true degrees of freedom of non-relativistic matter interacting with gravity at the weak field limit is of the form ${\cal H}_{mat}\otimes {\cal H}_{grav}$, where ${\cal H}_{mat}$ contains the matter degrees of freedom and ${\cal H}_{grav}$ the graviton degrees of freedom. If graviton degrees of freedom are frozen, or cast aside, as in the set up of the present paper and in Refs. \cite{Bose17, Vedral17}, all information about the  gravitational field is trivially contained in ${\cal H}_{mat}$. Any variability in the behavior of clocks in the spacetime is due to the quantum nature of the degrees of freedom in ${\cal H}_{mat}$ \cite{AnHu15}.

This is the key point where we  disagree not only with the claims of Refs. \cite{Bose17, Vedral17} but also with the arguments presented in Ref. \cite{ChRo} in support of them, about the quantum superpositions of geometries in Newtonian gravity. To two additional points made there: a) ``spacetime geometry can be in an entangled state with the particles", our response is, entanglement can only exist between two quantum objects. To make sense of the entanglement between geometry and particles one needs to show, not presume, the quantum nature of geometry or gravity.   b) ``Spacetime geometry is not just determined by the radiative degrees of freedom of gravity: it is also determined by the presence of matter." This is certainly true. In fact, we want to emphasize the correctness in the use of passive tense there for Newtonian force, which is completely slaved by the matter.  Only the  dynamical (radiative) degrees of freedom, the gravitons, can testify to the quantum nature of (perturbative) gravity.

\subsection{Comparison to the predictions of alternative quantum theories}

As explained in the introduction, our conclusions about the interactions between two gravcats were obtained under the assumption that the
usual quantization rules apply for matter in the presence of  gravity, in the weak-field, non-relativistic  (Newtonian) regime. This seemingly obvious assumption is not true in certain AQTs, like for example, the Diosi-Penrose scheme for gravitational decoherence. For this reason, we consider the prediction of specific AQTs for the set-up of two interacting gravcats of Sec. 2.

\bigskip

\noindent 1.  {\em DP-type gravity-induced decoherence}. In the Diosi-Penrose scheme, any cat state for a particle of mass $M$ and radius $R$ will be destroyed with a decoherence rate of $\frac{GM^2}{R}$. The decoherence rate grows with the size of the particle. For  an  optomechanical  nanosphere (see, e.g., Romero-Isart et al \cite{MechMQP,Pino}) with $M = 10^{10} amu$ and $R = 100 nm$, $\Gamma \sim 10^{-3}s^{-1}$. Note that the ratio $\Gamma/\mho \sim D/R >> 1$, where $D$ is the typical size of the gravcat.  Hence, in presence of the
 DP mechanism, no superpositions are possible in our proposed experiment, namely, no gravity-induced entanglement, no gravity-induced Rabi oscillations.

In terms of motivation, the proposers of the DP model seem to harbour the thought that gravity is not fundamentally quantum. However, this assumption is not needed in the derivation of the decoherence rate. %In this sense, we believe that the DP model is agnostic to the classical or quantum nature of gravity.
Nonetheless, its existence has important implications in the discussion about whether gravity should be quantized. For example,  Ref. \cite{BWGCBA} arguing that  (linearized)  gravity must be quantized  does not take the possibility of gravitational decoherence of the DP type into account.

Other models   that predict gravitational or fundamental decoherence in the position basis will generically predict that there are no  gravcats and no gravity-induced entanglement, or they will place stringent upper bounds to the gravcats' size and mass. Such models include (i) continuous collapse models like the Ghirardi-Rimini-Weber-Pearle models \cite{GRWreview} (ii) the Powers-Persival \cite{PowPer} collapse model from fluctuations of the conformal factor, and (iii) the
Asprea, Gasbari and Bassi model \cite{AGB} where  particles are under the influence of  a stochastic gravitational field.

\medskip

\noindent 2.{\em Newton-Schr\"odinger equation.} AQTs based on the Newton-Schr\"odinger equation (NSE) explicitly postulate that gravity is not quantum \cite{NS}.
In these AQTs, one  postulates a non-linear equation for the {\em single-particle} wave function,
\begin{eqnarray}
i\frac{{\partial}\psi}{{\partial t}}  = - \frac{1}{2m} \nabla^2 \psi + m^2 V_N[\psi]  \psi  \label{NS}
\end{eqnarray}
where %the potential energy $U(\bfr)$ from the gravitational interaction is of the form $U(\bfr)= m V_N(\bfr)$, where
$V_N(\bf r)$ is the (normalized) gravitational (Newtonian) potential given by
\begin{eqnarray}
V_N({\bf r},t) = - G \int d{\bf r'} \frac{|\psi({\bf r'},t)|^2 }{|{\bf r} - {\bf r'}|}.
\end{eqnarray}
 A NSE analysis of the two cat states leads again to a Hamiltonian of the form  (\ref{qubit}), for sufficiently small value of the particle's mass. It predicts no decoherence phenomena. Hence, the only difference is that the associated parameters, like $\mho$, are  defined with respect to solutions of the NSE, and not with respect to solutions of Schr\"odinger's equation. Hence, the NSE also predicts gravity-induced entanglement \footnote{Given a specific  potential,  the two-level approximation likely breaks down  for sufficiently large masses: if the Newtonian attraction  is sufficiently strong the lowest energy solution will not be concentrated around the two wells.  However, for any given mass, we can always choose a potential so that the Newtonian term will be significantly smaller than the height of the potential, so that the two level approximation is likely to be preserved. Hence, in principle, gravcats can form for any value of mass.}, in spite of postulating that gravity is fundamentally classical.

  %Note again that our results based on $N>> 1$ gives semiclassical gravity, not NS Equation. NSE for N equals to a few or a finite number, despite its name, cannot be derived from the nonrelativistic limit of quantum field theory plus the weak field limit of general relativity \cite{AHNSE}.

\medskip

\noindent 3. {\it ABH-type gravitational decoherence.}
The theoretical framework of the Anastopoulos-Hu \cite{AnHu13} and Blencowe \cite{Blen13} models of  gravitational decoherence is entirely within the confines of GR + QFT, without any AQT considerations.  The source of decoherence is the quantized weak perturbations off of the Minkowski spacetime. The source for Blencowe's decoherence is by thermal gravitons whereas Anastopoulos-Hu leave it open whether gravity is fundamental or emergent. Their difference shows up in a parameter $\Theta$ in AH's master equation corresponding to the noise temperature.  If gravity is fundamental then it is at the Planck temperature; if emergent it could have a broader range and at a lower temperature scale. The ABH master equations predict decoherence in the energy basis with rate $\Gamma \sim 10 \Theta (\Delta E)^2$, where $\Delta E$ is the difference in energy between two superimposed states. Position superpositions are not affected. The interaction of a pair of particles takes place through a Newtonian potential term in the Hamiltonian.

In this type of models, gravitational decoherence does not kill the gravcats.  For $\omega$ of the order of $eV$, and for $\Theta$ smaller than the Planck temperature, $\Gamma < 10^{-13}s^{-1}$.
Hence,  gravity-induced entanglement and Rabi oscillations are expected in the interaction of two gravcats, \textit{despite the fact that the theory treats the degrees of freedom of perturbative gravity as classicalized}.

\medskip

Our analysis demonstrates that gravity-induced entanglement is predicted also by theories that treat gravity (or linearized perturbations) as classical. This logically entails that \textit{the detection of gravity-induced entanglement is irrelevant to the quantum nature of gravity}, including that of linearized gravitational perturbations, and certainly a far cry from quantum gravity proper. As stated earlier, the Newtonian interaction is agnostic about the quantumness of gravity, because it does not involve the dynamical degrees of freedom of gravity.

\subsection{Foundational issues in gravitational effects of quantum matter}

The systems studied here are blind to the quantum or classical nature of gravity,  because we only deal with the Newtonian potential which is   slaved to the matter source. Our analysis only assumes that this property of the Newtonian potential passes on to the quantum theory.  Hence, the prime foundational question that can be settled by experiments invoking gravitational cat states is  the  following:  Does  the  gravitational  force  remain  slaved  to  the  mass  density  even  if the  latter  behaves  quantum  mechanically  (e.g., effects of  quantum fluctuations,  consequences of quantum measurements)?

 Unlike in Ref. \cite{AnHu15}, here we did not discuss the measurement of the gravitational force, we solely focused on identifying gravity-induced effects. A detailed analysis of such measurements is essential in order to elaborate on the relation between quantum observables and the classical spacetime picture of GR. To see this, we note that the gravitational potential in our system is an operator, that is subject to a probabilistic description. Classically, the potential is part of the spacetime metric that determines classical observables like   geodesic motion. A measurement scheme would determine a probability distribution for the geodesics followed by classical test bodies. Hence, fundamental issues like light-cone quantum fluctuations \cite{DeWitt, Isham, Ford} could be explored in a mathematically controlled and operationally well-defined context, that does not require a quantum gravity theory, and it is amenable to experimental testing.

 The cat states we consider here correspond to  the  two  minima  of  a  potential,  with  a  non-zero  tunneling  rate  between  them.    A recorded transition between
 two orthogonal qubit states is standardly described as a quantum jump. The familiar quantum jump experiments in atomic systems have shown that the duration of the jump is too small to be resolved, so jumps are effectively instantaneous.  This implies the  possibility  of  instantaneous  jumps  between  two  spacetime  geometries  that  correspond to  different  mass  distributions.  Newtonian gravity admits ``instantaneous” action, and in the non-relativistic regime one cannot explore issues of causality. So these simplifications from the full theories (GR + QFT) appear to be conveniently accommodating for the jump scenario.  Pushing this line further, the idea that quantum jumps can occur in the spacetime description because of the interaction of gravity with quantum matter is a novel phenomenon, yet of  foundational value, worthy of closer experimental attention.

 \end{document}